\documentstyle[aps,pre,epsfig,graphicx]{revtex}

\newcommand {\bc} {\begin{center}}
\newcommand {\ec} {\end{center}}
\newcommand {\bd}{\begin{displaymath}}
\newcommand {\ed}{\end{displaymath}}
 \newcommand {\be} {\begin{equation}}
\newcommand {\bea} {\begin{eqnarray} \nonumber }
\newcommand {\ee} {\end{equation}}
\newcommand {\eea} {\end{eqnarray}}
 \newcommand {\eps} {\epsilon}

 \newcommand {\al} {\alpha}

\def\eps{\epsilon}
\def\al{\alpha}

 \def\(({\left(}
 \def\)){\right)}
\def\[[{\left[}
\def\]]{\right]}
\def\bi{\bibitem}

\def \form#1 {eq. (\ref{#1}) }
\def \parziale#1#2  {{\partial {#1} \over \partial {#2}}}

\begin{document}
\title{Statistical Physics of the  Glass Phase}
\author{Marc M\'ezard}
\address{Laboratoire de Physique Th\'eorique et Mod\`eles Statistiques
\\
Universit\'e Paris Sud, Bat. 100, 91405 Orsay {\sc cedex}, France \\
mezard@ipno.in2p3.fr}

\date{\today}
\maketitle

\begin{abstract}
This paper, prepared for the
proceedings of the 'Statphys 21' conference (Cancun, july 2001)
gives an introduction to the statistical physics problems
which appear in the study of structural glasses.
It is a  shortened and updated version of a more detailed review paper 
 which has appeared in \cite{mmaspen}.
\end{abstract}

Under a rapid enough cooling,
 almost any  liquid becomes a glass\cite{glass_revue}. This is
a state of matter in which 
the density profile, as observed on accessible time scales,
is not flat as in a liquid, it contains some
 peaks as in a crystal, but these peaks are not located on the nodes
 of a periodic or quasi periodic lattice. Particles tend to localize around
some fixed positions, which appear random.

Glasses have been recognised for their technological 
importance for several millenia, and a large proportion of matter in nature is
found in a glass state. But they also offer a wonderful set of 
theoretical questions,
most of which are very far from being understood.

The first such question which comes to mind is whether  the glass is a new state 
of matter, or whether it is just a liquid state where the relaxation
to uniform density has become so slow that it is inobservable.
The question is difficult because the glass state
 is not distinguished by any obvious symmetry (a not-obvious 
symmetry will be discussed later) from the liquid state.
A similar problem was studied in the context  of spin glasses, which
also undergo a transition from a paramagnetic 'liquid' state to a state
where the spins are frozen in random directions. In  
spin glasses, experiments have provided convincing evidence for the
existence, at least in zero magnetic field,
 of a critical temperature where the nonlinear part of the
magnetic susceptibility diverges with well defined critical exponents.
Thus the experiments on spin glasses tend to favour the existence of a phase
transition and of a new spin-glassy state of matter \cite{youngbook}. This behaviour
is also found in the study of mean field theories\cite{MPV}, and recent simulations
on some model three dimensional spin glasses also point in this same direction,
but there does not exist any proof of the 
transition in any finite dimensional spin glass model: 
a challenge to mathematical physics!

Similar challenges, but much more difficult, exist in the study of structural
glasses\cite{kepler}, and one may fear that relevant rigorous results will
not be found in the near future.
Let us examine how one can formulate the problems
of  structural glasses from a restricted statistical physics point of view.
 One wants to start from 
a microscopic Hamiltonian. The simplest situation is that of  $N$ point-like
particles in a volume $V$:
\be
H= \sum_{i<j} V(r_i-r_j) \ ,
\ee
 with a pair interaction potential$V(r)$ which
 can be for instance either a hard sphere potential,
a `soft sphere' potential ($V(r)=A/r^{12}$), or a Lennard-Jones
 potential ($V(r)=A/r^{12}-B/r^6$). Because these systems
tends to crystallize even on the short time scales accessible numerically,
one often uses\cite{kobrev} the binary mixtures where there are two types
of particles $A,B$, and three different interaction
potentials $V_{AA}(r),V_{BB}(r),V_{AB}(r)$.

Here is a small sample of questions one would
like to ask on these systems, from a statistical physics point of view.
\begin{itemize}
\item {\it ``Equilibrium''  questions}:

Given the interaction potential, is there a true phase transition
to a new state of matter? Is there a transition nearby in parameter space, 
which is never seen, for instance because it is in an unphysical region of
parameters, or because the time scales make it impossible
to reach the transition? 
Does there exist some  choice of the V's such that  
the crystal is frustrated enough, in such a way that 
 the true equilibrium state below some temperature is actually a
 glass state?  Does there  exist at least some choices
where the glass state exists as a long-lived metastable state (like
the diamond phase of carbon)?
Is there a generic geometric description of the glass transition
 in terms of potential energy landscape,
or of free energy landscape? What is the role of geometric frustration,
the role of dimension?
\item {\it ``Dynamical'' questions}:

Do these ``equilibrium''  questions (or more precisely the answers to them)
 have anything to do with the experimentally observed processes of 
extreme slowing down upon decrease in temperature, or with the
out of equilibrium dynamics observed in the glass phase? 
Are there more universal descriptions of systems falling out of
equilibrium, which are common to some systems having a glass transition
and others being always (slow) liquids?
How does the choice of the interaction potential, and the corresponding 
energy landscape, affect the dynamics and
the glass transition?
\end{itemize}
These questions are just meant to point out that glasses 
offer some basic challenges to statistical physics, on top of
all the challenges they also offer from different perspectives
like chemistry and material sciences.

One should not forget either that some type of glassy states are often encountered
in many different fields outside of physics, and progress in the study
of glass phases has already brought new results 
in many different areas such as
combinatorial optimization\cite{MPV,optrev}, error correcting codes \cite{codes}, 
or  behaviours of interacting agents \cite{mingame}. 
In fact one often encounters in these
fields some problems which are simpler than those of the glass phases
found in nature, because there is no spatial structure: the interaction between
various 'atoms' (constraints, agents,..) typically takes place between randomly
chosen atoms, without a constraint of spatial neighborhood. This allows
for some exact solutions using mean field like methods.

Turning back to structural glasses,
even if one forgets about rigorous mathematical proofs, I must say that all 
of the above questions are unanswered: there are some hints of answers
coming from experiments, simulations, theories on simple systems, 
mean field theories, etc..,
and the progress in the last few years is real, but there is no real consensus
around one single and  consistent scenario. Let us see some pieces of this puzzle.
	
Experimentally, when crystallisation is avoided,
a supercooled liquid falls out of equilibrium on experimental 
time scales \cite{fn1}, and 
becomes a `glass',
at a temperature $T_g$ called the glass temperature\cite{glass_revue}. This 
glass temperature is conventionnally defined as the one at 
which the relaxation time $\tau$ of the liquid, as obtained e.g. from
viscosity or from susceptibility measurements,
becomes of the order of $10^3$ seconds. 
Angell's plot of 
$\log\(( \tau /1 s \))$ versus $T_g/T$ allows to distinguish 
several types of behaviour.
So called strong glasses like $SiO_2$ have 
$\log\(( \tau /1 s \)) =A (T_g/T)$,
a typical Arrhenius behaviour 
with one well defined free energy barrier. 
On the other hand, some glasses, called fragile, 
show a dramatic increase of the relaxation time
when decreasing temperature which is much faster than Arrhenius:
the typical free energy barrier thus increases when $T$ decreases.

A popular fit of the relaxation time versus temperature is the Vogel Fulcher one,
\be
\tau \sim \tau_0 \exp\(({A \over T-T_{VF}}\))
\ee
which would predict a phase transition at a temperature
$T_{VF}$ which is not accessible experimentally (while staying at equilibrium).
The more fragile the glass, the closer is $T_{VF}$ to $T_g$, while strong glass formers have a
$T_{VF}$ close to zero.
 The increase of barrier heights when $T$ decreases in  fragile glasses
might imply a collective behaviour involving more and more particles,
however no sign of a divergent static correlation length has been found so far
(an increasing dynamic length has been found recenty \cite{donati}).

Another much studied property is  the specific heat.
When one cools the liquid slowly, at a cooling rate $\Gamma=-dT/dt$, it 
freezes into a glass at a temperature
which decreases slightly when $\Gamma$ decreases. 
When this freezing occurs, the specific heat
jumps downward, from its value in the equilibrated supercooled liquid
\cite{fn1} state to 
a glass value which is close to that of the crystal. This is a signature of
the fact that the system below the freezing temperature 
is non ergodic on experimental time scales. 
From the specific heat, one can compute  configurational entropy, defined 
 as the difference $S_c=S_{liq}-S_{crystal}$. It
decreases  smoothly with $T$ in the supercooled liquid phase, until the system 
becomes a glass. It was noted by Kauzmann long ago that, 
if extrapolated, $S_c(T)$ vanishes at
a finite temperature $T_K$. If cooled more slowly, the system
follows the smooth $S_c(T)$ curve down to slightly lower temperatures, 
but then freezes again.
 One can wonder what could happen at infinitely slow cooling. As a
negative $S_c$ does not make sense 
(except for pure hard spheres, where there is no energy),
 something must happen at temperatures above $T_K$. 
The curve $S_c(T) $ could flatten down smoothly,
or there might be a phase transition, which in the simplest
scenario would lead to $S_c(T)=0$ at $T<T_K$. This idea of an underlying "ideal"
phase transition \cite{fn2}, which
could be obtained only at infinitely slow cooling, receives some support from
the following observation: 
the two temperatures where the {\it extrapolated} experimental 
behaviour has a singularity,
$T_{VF}$ and $T_K$, turn out to be amazingly close to each other 
\cite{RiAn}.
The first phenomenological attempts to explain this fact 
 originate in the work of Kauzmann \cite{kauzmann}, and 
developed among others 
by Adam, Gibbs and Di-Marzio \cite{AdGibbs}, which identifies the glass transition 
as a `bona fide'
thermodynamic transition blurred by some dynamical effects.

If there exists a true thermodynamic glass transition at $T=T_K=T_{VF}$,
it is a transition of a strange type.
It is of second order because the entropy and internal energy are 
continuous. 
 On the other hand the order 
parameter is
discontinuous at the transition, as in first order transitions: the
modulation of the microscopic density profile in the glass does not appear continuously from the
flat profile of the liquid. As soon as the system freezes, there is a finite jump
in this modulation 
(A more precise definition of the order parameter will be given below).

It turns out that  such a 1st-2nd order type transition has been found in 
the theoretical study of some mean field models for a 
certain category of spin glasses.
A few years after  the replica symmetry breaking (RSB) 
solution of the mean field theory of spin glasses \cite{MPV}, it was 
realized 
that there exists   another category of mean-field spin
glasses where the static phase transition exists 
and is due to an entropy crisis \cite{REM}. These
are now called discontinuous spin glasses because their phase transition, 
although
it is of second order in the Ehrenfest sense, has a discontinuous order
parameter \cite{GrossMez}.
An alternative name is 
 `one step RSB' spin glasses (because of the
special pattern of symmetry breaking involved in their solution)
and the new type of transition is sometimes named a 'random first order transition'.
 
The archetypes of discontinuous  spin glasses involve
 infinite range interactions between triplets (or higher order
groups) of spins.
The simplest among them is the random energy model, which is the $p \to \infty$ limit
version of the p-spin models described by the Hamiltonian 
\be
H=-\sum_{i1<...<i_p} J_{i_1...i_p}
s_{i_1}...s_{i_p}
\label{hamil}
\ee
 where the $J$'s are (appropriately scaled) quenched random couplings,
and the spins can be either of Ising or spherical type \cite{GrossMez,KiThWo,crisanti}. The analogy
between the phase transition of discontinuous spin glasses and the 
 'transition' in structural glasses  was first noticed 
in the mid-eighties \cite{KiThWo}. While some of 
the basic ideas of the present development were around at that time, 
there still missed a few crucial ingredients. 

One big obstacle was the existence (in spin glasses) versus
the absence (in structural glasses)  of quenched disorder. The 
discovery  of discontinuous spin glasses without any 
quenched disorder
\cite{nodis1,nodis2,nodis3}
provided an important new piece of information: contrarily to what had 
been
believed for long, quenched disorder is not necessary for the existence of
a spin glass phase. In fact it has been found recently that 
some suitably defined systems can self induce disorder and frustration
in their dynamical behaviour, even if these properties
are not present in the system a priori. 

\begin{figure}[bt]
\centerline{    \epsfysize=6cm
       \epsffile{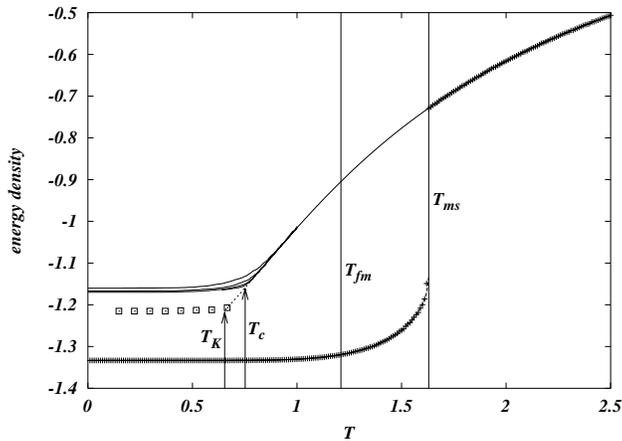}}
\caption{Average energy as a function of the temperature for the
ferromagnetic triangular plaquette model, on a Bethe
hyperlattice where each spin belongs to  $k=4$ plaquettes 
(from[22]).
If one prepares the system at low temperature in its ferromagnetic
configuration $s_i=1$, and heats it up, the energy follows the bottom
curve. At temperature $T_{fm}$ the ferromagnetic state becomes metastable,
but the system stays in this metastable ferromagnetic state until 
the spinodal temperature $T_{ms}$ is reached. It then becomes paramagnetic.
Cooling down from the paramagnetic state, the energy follows the upper curve,
 where the magnetization is zero. Below $T_{fm}$ one is in a regime of supercooled 
paramagnetic state.
In the temperature range below $.9$, 
one sees the effects of the cooling rate dependance on the energy (cooling
rates from $10^2$ to $10^5$ MCS per $\Delta T=0.01$). The theory predicts
that at infinitely slow cooling there is a dynamical transition at $T_c$,
where the energy is blocked at a threshold value, while the equilibrium 
ideal glass transition takes
place at $T_K$, and the equilibrium energy density below $T_K$ is given by
the squares. It is conjectured that in finite dimensional models the
temperature $T_c$ marks the onset of activated processes: the energy
can decrease slowly, and the real transition takes place at $T_K$.}
\label{ferropspin_fig}
\end{figure}

Here I would like to mention one such  system, which displays some 
properties strikingly similar to structural glasses, and can be 
studied in great details.
This example is a 
purely ferromagnetic $p=3$
spin model of the type (\ref{hamil}). The spins $s_1,s_2,s_3$
around a plaquette interact
with a ferromagnetic term $-J s_1 s_2 s_3$. If the plaquettes are organised
in a Bethe lattice of plaquettes
where each spin belongs to exactly $k$ plaquettes, 
one has no disorder on any finite lengthscale, and no frustration (the
ground state is unique and purely ferromagnetic: all spins up).
One can in fact solve this problem completely
\cite{ferropspin}. There exists a ferromagnetic phase, 
but it is very difficult to find it
dynamically, and instead the spins tend to freeze into a glass phase
(see fig. \ref{ferropspin_fig}). The reason is easy to understand:
if the spin $s_1$ has a wrong orientation ($s_1=-1$), the
plaquette interaction induces an effective {\it antiferromagnetic} interation
betwen $s_2$ and $s_3$, and the system is thus frustrated, and disordered,
unless it has found its exact ferromagnetic ground state $s_i=1$.
Similar interesting 'plaquette' model have been described in three
dimensions, and studied numerically\cite{plaq}.

\newpage 

\begin{figure}
\hbox{ \includegraphics[width=0.4\textwidth,angle=-90]{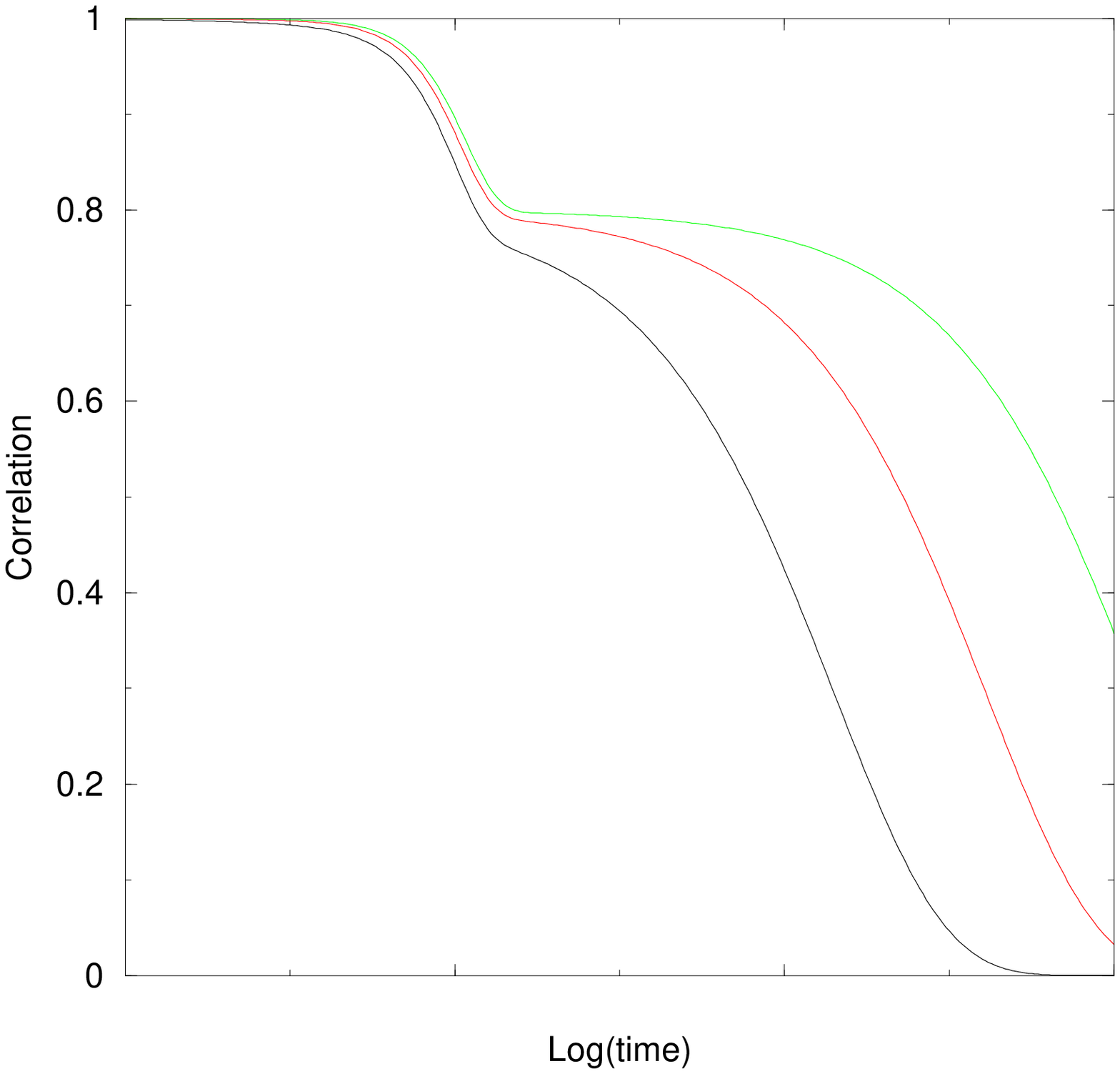}
 \includegraphics[width=0.4\textwidth,angle=-90]{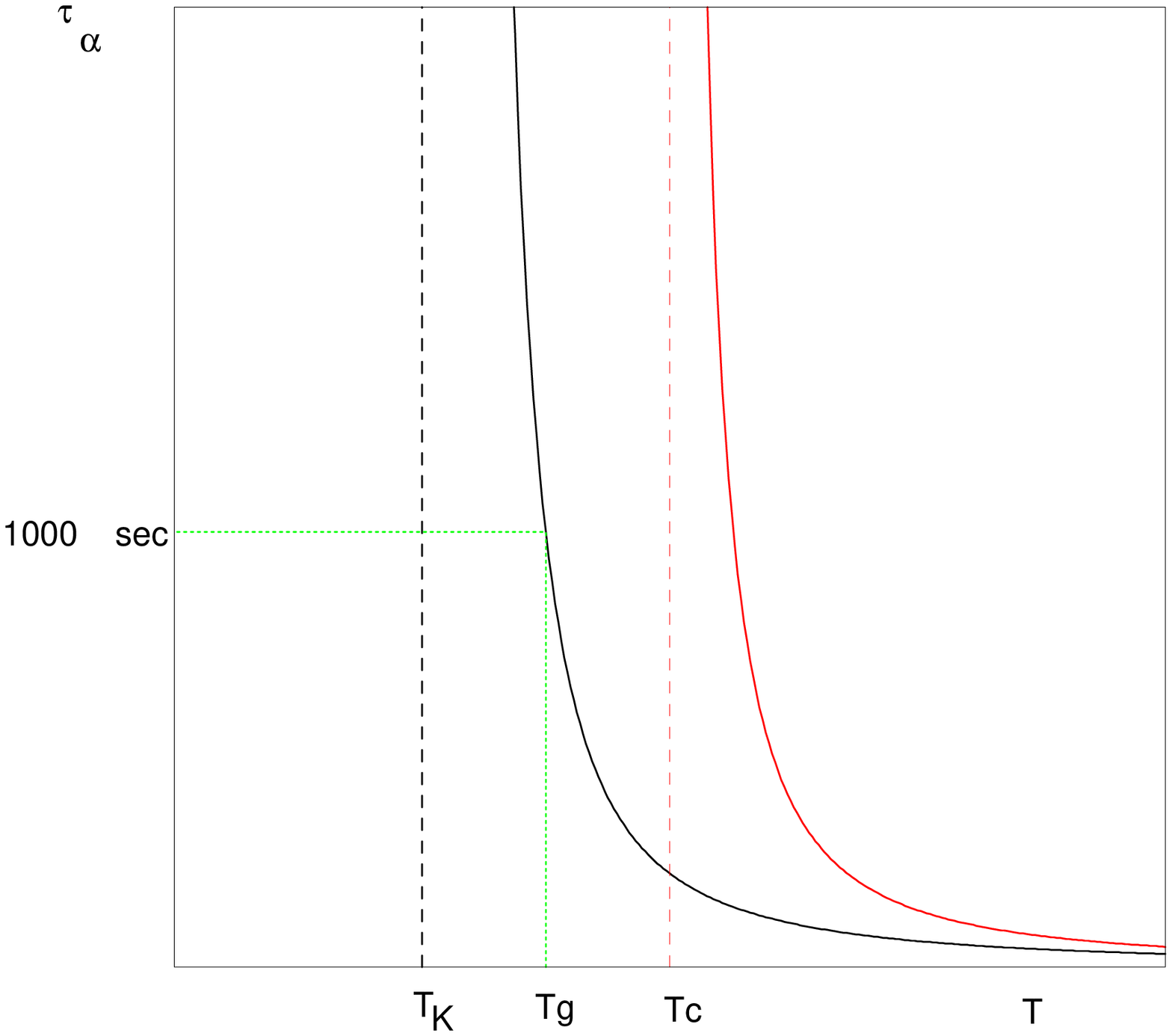}}
%
\caption{The left-hand figure shows the schematic behaviour of the correlation function found in
mean field discontinuous spin glasses and observed in structural glasses.
The typical two-step relaxation consists of  a
fast $\beta$ relaxation leading to a plateau, followed by a
$\alpha$ relaxation from the
plateau, whose typical time scale  increases rapidly when $T$ decreases,
and diverges at $T=T_c$ which is equal to the
mode-coupling transition temperature. 
The right-hand figure shows the behaviour of the 
relaxation time versus temperature. The right-hand curve is the prediction
of mode-coupling theory without any activated processes: 
it is a mean field prediction,
which is exact for instance in
 the discontinuous mean-field spin glasses. 
The left-hand curve is the observed
relaxation time in a glass. The mode coupling theory 
provides a quantitative prediction for the increase
of the relaxation time when decreasing temperature, 
at high enough temperature (well above the
mode coupling transition $T_c$)
[24,25]. 
The departure from the mean field prediction at lower temperatures is 
usually attributed to 'hopping' or 'activated' processes, in which the system is trapped
for a long time in some valleys, but can eventually jump out of it. The ideal
glass transition, which  takes place at $T_K$, 
cannot be observed directly since the system 
becomes out of equilibrium on laboratory time scales at the `glass temperature' $T_g$.
 }
\label{twostep}
\end{figure}

A closer look at the solution of a discontinuous spin glass problem
such as the one in fig.\ref{ferropspin_fig} shows that there are actually
two transition temperatures. There is a true equilibrium transition
at a temperature $T_K$, with a jump downward of the specific heat.
However there also exists a dynamical transition temperature 
at a temperature $T_c$ which is larger than the equilibrium transition $T_K$. 
When T decreases and gets near to $T_c$,
the correlation function relaxes with a characteristic two steps form: a
fast $\beta$ relaxation leading to a plateau takes place on a characteristic time
which does not grow, while the $\alpha$ relaxation from the
plateau takes place on a time scale which diverges when $T \to T_c$ (see fig. \ref{twostep}). 
This dynamic
transition is exactly described by the schematic mode coupling
 equations\cite{gotze,MCexp}. 
The presence of this two steps relaxation is a well known experimental fact 
in structural glasse, and a lot of  work has shown that many 
observed details of
this two times relaxation are well characterized by the mode coupling theory,
provided one does not get too close to $T_c$. However,  mode coupling theory
would predict that the relaxation times diverges at $T_c$, where experiments
clearly find a finite relaxation time. From the point of view of
discontinuous spin glasses, the
existence of a dynamic transition is associated wiht the fact that
the system can be trapped in metastable states with an extensive excitation
free energy above the equilibrium state. This
is possible only in mean field.  In finite dimensions, nucleation effects
(called activated effects in the mode coupling litterature)
prohibit such a situation. An important  conjecture\cite{KiThWo}
 is that in a realistic system like a glass, 
the region between $T_K$ and $T_c$ will have instead a finite, but very rapidly 
increasing, relaxation time, as explained in fig. \ref{twostep}. A similar 
behaviour has been found in finite-size mean field models \cite{cririt}.

 Another very interesting dynamical regime is the  out of equilibrium one
($T<T_g$). Then the system is no longer stationnary:
it ages. Schematically, new
relaxation  processes
come into play on a time scale comparable to the age of the system: the older
the system, the longer the time needed for this "aging" relaxation to take place.
Recent years have seen  important developments
 on the out of equilibrium dynamics in glassy phases \cite{BCKM},
initiated by the exact solution of the dynamics in a discontinuous spin 
glass \cite{cuku}. It has become clear that, in realistic systems 
with short
range interactions, the pattern of replica symmetry breaking can be 
deduced
from the measurements of the violation of the fluctuation dissipation 
theorem \cite{fdr}.
These  measurements are difficult. However, numerical
simulations performed on different types of glass forming
systems have provided an independent and spectacular confirmation of their
`one step RSB' structure \cite{gpglass,bk1,bk2,leo} on the (short) time
scales which are accessible.
Experimental results have not yet settled the issue, but the first measurements of
effective temperatures in the fluctuation dissipation relation have been made  
recently \cite{fdr_exp}. One should also underline that the concepts of
effective temperature goes beyond the restricted field of glasses and
is being currently tested on various problems like granular media \cite{jorge}.

To summarize, the analogy between the phenomenology of fragile glass formers and 
discontinuous mean field spin glasses accounts for:
\begin{itemize}
\item The discontinuity of the order parameter
\item The continuity of the energy and the entropy
\item The jump in specific heat (and the sign of the jump)
\item Some kind of "entropy crisis"  \`a la Kaufmann (see below)
\item The two steps relaxation of the dynamics and the success of Mode Coupling 
Theory at relatively high temperatures.
\item
The aging phenomenon and the pattern of modification 
of the fluctuation dissipation relation in the low temperature phase.
\end{itemize}

\begin{figure}[bt]
\centerline{    \epsfysize=6cm
       \epsffile{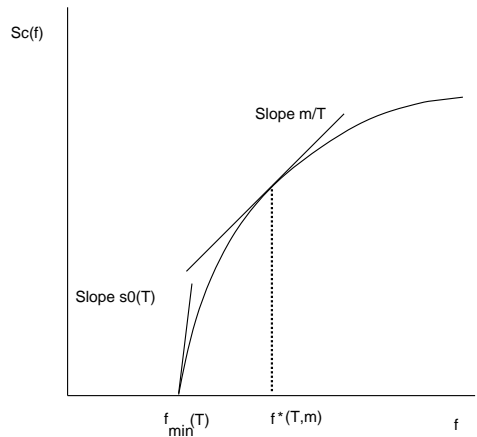}}
\caption{
Qualitative shape of the complexity versus versus free energy
in  mean field discontinuous spin glasses. 
The whole curve depends on the temperature. The
saddle point which dominates the partition function,
for $m$ constrained replicas, is the point $f^*$ such
that the slope of the curve equals $m/T$ (for the usual unreplicated system, 
$m=1$).
If the temperature is small enough the saddle point sticks
to the minimum $f=f_{min}$ and the system is in its glass phase.
For $m=1$, this equilibrium phase transition happens at $T=T_K$.  }
\label{sigma_qualit}
\end{figure}

Because of the successes of the above analogy, it is worth to
study the basic ingredients at work in the glass transition of the 
mean field discontinuous spin glasses.
At temperatures $T_K<T<T_c$, 
the phase space breaks up into
ergodic components which are well separated, so-called free energy valleys or TAP
states \cite{crisomtap}. This has been confirmed recently
by some rigorous implementation of the cavity method \cite{talag}. 
Each valley $\al$ has a free energy $F_\al$ and a free energy 
density 
$f_\al= F_\al/N$. The number of free energy minima with 
free energy density  $f$ is found to be exponentially large:
\be
{\cal N}(f,T,N) \approx \exp(N\Sigma(f,T)),\label{CON}
\ee
where the function $\Sigma$ is called the complexity. 
The total free energy of the system, $\Phi$, 
can be well approximated by:
\be
 e^{-\beta N \Phi} \simeq \sum_\al e^{-\beta N f_\al(T)} =
\int_{f_{min}}^{f_{max}} df \ \exp\((N[ \Sigma(f,T)-\beta f]\)) \ ,
\label{SUM}
\ee
where $\beta=1/T$.
The minima which dominate the
sum are those with a free energy density $f^*$
 which minimizes the quantity $\Phi(f)=f-T\Sigma(f,T)$.
At large enough temperatures the saddle point is at $f>f_{min}(T)$. When 
one
decreases $T$ the saddle point free energy decreases (see fig.\ref{sigma_qualit},
with $m=1$).
The Kauzman temperature $T_K$ is that below which the saddle point sticks
to the minimum: $f^*=f_{min}(T)$. It is a genuine phase transition \cite{REM,GrossMez,KiThWo}.
However because $T_c>T_K$, the phase space is actually separated into 
non ergodic components (valleys) at $T<T_c$. The total 
equilibrium free energy is analytic at $T_c$: in spite of the
ergodicity breaking, the system has the same free energy as that of the liquid,
as if transitions were allowed between valleys.

What remains of this mean field picture in finite dimensional glasses?
When one decreases the temperature, there is a well defined separation of 
time scales between the $\alpha$ and the $\beta$ relaxations, which suggests to
consider the dynamical evolution of system
in phase space  as a superposition of two processes: an intravalley (relatively fast)
relaxation, and an intervalley (slow, ang getting rapidly much slower when one cools the system)
hopping process. 
One popular way of making this statement more precise,
allowing  to study it numerically, is through
the use of inherent structures (IS) \cite{IS,inherent}. Given a configuration of
the system, characterized by its phase space 
position $x=\{ {\vec x_1},...,{\vec x_N}\}$, 
the corresponding inherent structure $s(x)$ is another 
point of phase space which is the
local minimum of the Hamiltonian which is reached from 
this configuration through a 
steepest descent dynamics. IS are easily identified numerically \cite{schroder}.
The number of IS at a given energy is often also 
called the configurational entropy.

Should one identify the IS with the free-energy valleys, 
and the configurational entropy with the complexity? The answer is no.
The difference is
very easily seen in spin systems \cite{BiMo}: IS are nothing but configurations
which are stable to one spin flip. 
Zero temperature free energy valleys, defined as TAP states,
are stable to the flip of  any arbitrarily large number $k$ of spins (but the
limit $N \to \infty$ must be taken before the limit $k \to \infty$). In continuous
systems, the generalization is clear: 
IS are local minima of the energy, so that any infinitesimal
move of the positions of all $N$ particles raises the energy. Let us
generalize the notion of a minimum as follows: define a k-th order local minimum as
a configuration of particles such that any infinitesimal move of all $N$ particles, 
together with a move of {\it arbitrary size} of $k$ particles, raises the energy. 
The limit
$k \to \infty$ gives the  definition of a zero temperature ``free'' energy valley.
The proper definition at finite temperature is known but
is slightly more involved  \cite{pot} and I shall briefly allude to it
below, after defining clones.
Note that a zero temperature free energy valley cannot be identified 
by the fact that the Hessian is positive definite: it is necessary that the
Hessian be positive in such a state, but this is not a sufficient condition.

The 'entropy crisis scenario' which associates the ideal glass transition
with the vanishing of some type of configurational entropy,
should thus be taken with a grain of salt in a finite dimensional system.
If one considers valleys as IS, and the corresponding configurational entropy,
in a system with short range interactions, it is reasonable to expect that 
one may have two distinct IS which differ by a local rearrangement of
a finite number of atoms. 
It is then easy to show that the slope of configurational
entropy versus free energy is infinite around $f_{min}$ \cite{Still_slope}, which does not
agree with the general scenario, except if the Kauzmann temperature vanishes.
If one considers valleys as free energy valleys, then
 two configurations which differ by the (arbitrarily large) displacement
of a finite number of atoms are in the same thermodynamic valley. Nucleation
arguments then forbid the existence of a non-trivial
complexity versus free energy curve in a finite dimensional system: 
one cannot have excited free energy valleys with an 
extensive excitation free energy.
The solution consists in noticing that
there do exist many metastable valleys, with a finite complexity,
but these valleys have   a finite but very long lifetime
(this discussion is very similar to that of the metastability of the
supercooled liquid). If this lifetime is much larger than experimental times,
the 'mean field' like description neglecting nucleation is a valid approach. 

Having somewhat criticized the idea of inherent structure, I would like
to point out that the energy landscape is nevertheless a useful tool,
not only because it is simpler to describe than the true free energy valleys.
Recent studies of the 
energy landscape in the supercooled
liquid phase have shown some evidence that the typical
energy $E_c$ of IS found at $T_c$ marks 
a kind of topological transition between a phase where the  nearest saddle of
a generic configuration has some unstable directions (above $E_c$),
and another phase (below $E_c$) where it is a local minimum \cite{landscape}.
This allows for an interesting conjecture on
 the difference between fragile and strong glass formers
\cite{cav}: when the energy decreases through $E_c$, the
dynamics is necessarily activated; if the barriers found at $E_c$ are large
compared to $k T_c$, the system is fragile, otherwise it is strong.

Let us now see briefly how one can go beyond the simple analogy 
between structural glasses and mean field discontinuous 
spin glasses. One can actually use the concepts and the techniques which
are suggested by this analogy in order to 
start a systematic first principles study of the
glass phase \cite{MePa1,MePa2,MP_Trieste}.  
The idea is to postulate the existence of a scenario
for the glass transition such as the one found in discontinuous 
spin glasses, and see how one can use it for computing 
properties of structural glasses. The validity of such an approach is validated by
the predictions it can make. It is not able to prove the existence
of a transition. However in physics one often uses
very successfully this kind of
reasoning, and one should not focus too much on the
absence of an existence proof (nobody has proven to you the existence of crystals 
during your solid state physics course). In fact it could even be that there is no
ideal phase transition, the metastable states always have a finite lifetime,
but this is so large at low temperatures (around $T_K$)
 that the present postulate is a very good starting point 
(as when one postulates the existence of diamond).
A severe limitation at the present stage of development of the theory is
that, for technical reasons, we have been able to implement this program only
for computing equilibrium properties: we don't know yet how to
perform a first principle dynamical computation going
in the out of equilibrium regime, below
the mode coupling temperature (see however \cite{latz}). Instead we shall compute equilibrium properties
(typically $T_K$, the radius of the cage which confine 
the particles in 
the glass phase, the configurational entropy etc... )
which can be confronted to smart computer 
simulations on small systems (see e.g. \cite{krauth,parisi_c}.

The first task is to define an order parameter. This is not trivial in an 
equilibrium theory where
we have no notion of time persistent correlations. The
problem is that  the glass state seems to have the same symmetry 
as that of a liquid.
However there is a kind of symmetry which is broken in the glass state,
and it can be identified by looking at correlations between
two replicas of the systems. One  introduces two sets of particles
which have
positions $x_i$ and $y_i$ respectively, the total Hamiltonian is
\be
E=\sum_{1 \le i \leq j \le N}( v(x_i-x_j) +v(y_i-y_j))+\eps \sum_{i,j} 
w(x_i-y_j)
\label{couplage}
\ee
where we have introduced a small attractive potential $w(r)$ between
the two systems. The precise shape of $w$ is irrelevant, insofar as we shall
be interested
in the limit $\eps \to 0$, but its range should be of order or smaller than
the typical interparticle distance. The order parameter is 
then the correlation function between the two 
systems:
\be
g_{xy}(r)=\lim_{\eps \to 0} \lim_{N \to \infty} 
{1 \over \rho N} \sum_{ij} <\delta(x_i-y_j-r)>
\ee
In the liquid phase this correlation function is identically equal to one,
while 
it
has a nontrivial structure in the glass phase, reminiscent of the pair 
correlation
of a dense liquid, but with an extra peak around $r \simeq 0$. 
The idea behind this order parameter is that, if there exist favoured 
glass structures, they are very difficult to determine a priori, 
and we don't know how to introduce an external field in order to 
polarize the system into one of them. However, the second copy, 
with a weak attraction to the first one, just plays this role
of an infinitesimal external field, necessary in order to study a symmetry 
breaking process (here the symmetry is the global
translation of the $y$ particles with respect to the $x$ particles). 
Let us notice that we
expect a discontinuous jump of this order parameter at the transition, in spite
of the fact that the transition is of second order in the 
thermodynamic sense.

Generalizing this approach to a system of $m$ coupled replicas,
sometimes named `clones' in
this context (the order parameter used only $m=2$), provides a wonderful
method for studying  analytically the thermodynamics of the glass phase
\cite{remi,Me}.  In the glass phase, the 
attraction will force all $m$ systems
 to fall into the same glass state, so that
the  partition function is:
\be
Z_{m} = \sum_\al e^{-\beta Nm f_\al(T)}= \int_{f_{min}}^{f_{max}} df
 \ \exp\((N [ \Sigma(f,T)-m \beta f]\))
\label{zm}
\ee
In the limit where $m \to 1$ the corresponding partition function 
$Z_m$ is dominated by the correct saddle point $f^*$ for $T>T_K$. 
The interesting regime is when the temperature is  $T<T_K$, 
and the number $m$ is allowed to become smaller than one. The saddle 
point $f^*(m,T)$ in the expression (\ref{zm}) is the solution
of $\partial \Sigma(f,T) / \partial f=m/T$. Because of the
convexity of $\Sigma$ as function of $f$, the saddle point is
at $f>f_{min}(T)$ when $m$ is small enough, and it   sticks at 
$f^*=f_{min}(T)$
when $m$ becomes larger than a certain value $m=m^{*}(T)$,
a value which is smaller than one when $T<T_K$ (see fig. \ref{sigma_qualit}). 
The free energy
in the glass phase, $F(m=1,T)$, is equal to $ F(m^*(T),T)$. As the free 
energy
is continuous along the transition line $m=m^*(T)$, one can compute 
$F(m^*(T),T)$ from the region $m \le m^*(T)$, which is a region where the
replicated system is in the liquid phase. This is the clue to
the explicit computation of the free energy in the glass phase. 
It may sound a bit strange because one is tempted to think of $m$ as an 
integer
number. However the computation is much clearer if one sees $m$ as 
a real parameter in (\ref{zm}). As one considers low temperatures $T<T_K$ 
the
$m$ coupled replicas fall into the same glass state and thus they build
some molecules of $m$ atoms, each molecule being built from one atom of 
each 
'colour'. Now  the interaction strength of one such molecule with another 
one
is basically  rescaled by a factor $m$ (this 
statement becomes  exact in the limit of zero temperature
where the molecules become point like). If $m$ is small enough this 
interaction is small
 and the system of molecules is liquid. When $m$ increases, the molecular 
fluid
freezes into a glass state at the value $m=m^*(T)$.
So our method requires to estimate the 
 replicated free energy, 
$
F(m,T)=-{\log(Z_m) /( \beta m N )}
$,
 in a molecular
liquid phase, where the molecules consist of $m$ atoms and
$m$ is smaller than one. For $T<T_K$, $F(m,T)$ is maximum at
the value of $m=m^{*}$ smaller than one,
while for $T>T_K$ the maximum is reached at a  value $m^*$  larger than one.
  The knowledge of $F_m$ as a 
function
of $m$ allows to reconstruct the configurational entropy
function $Sc(f)$ at a given temperature $T$
through a Legendre transform.
 The Kauzmann temperature ('ideal 
glass
temperature') is the one such that $m^*(T_K)=1$. For $T<T_K$ the equilibrium
configurational entropy vanishes.
This gives the main idea which allows to compute the free energy in the
glass phase, at a temperature
$T<T_K$, from first principles: it is equal to the free energy of a molecular liquid
at the same temperature, where each molecule is built of $m$ atoms, and
an appropriate analytic continuation to $m=m^*(T)<1$ has been taken. The whole problem 
is reduced to a computation in a liquid. This is not trivial, and requires to develop
some specific approximations. I shall not elaborate on that here, but 
refer the reader to the original papers \cite{MePa2,sferesoft,LJ,LJ2}.
Let me just mention one point which one should keep in mind: so far this analytic
approach involves some low temperature expansion, where the possibility for
two molecules to exchange atoms is not taken into account. 
Within this approximation the
difference between pure states and IS is not seen.
The results are in good agreement with numerical simulations on various systems like
binary soft spheres or binary Lennard-Jones; a good test is the computation of the configurational entropy. One should also notice that recent numerical simulations
on small systems point towards the possible existence of an equilibrium phase transition at a temperature $T_K$ close to that predicted by this theory 
\cite{parisi_c}(to be precise
the theory assumes the existence of this ideal glass transition, and can then compute the value of $T_K$).

The existence of metastable states in the temperature range $T_K,T<T_c$ 
can be confirmed by studying the potential $W(q)$ which is the 
 Legendre transform of the free energy $F(\eps)$ for one replica (particles $y$)
coupled as in (\ref{couplage}) to a reference equilibrated system (particles $x$)
\be
W(q)=F(\eps)+\eps q \ \ \ ; \ \ \ q={-\partial F \over \partial \eps} \ .
\ee
Analytic computation in mean field models \cite{pot}, as well as in  glass forming liquids 
using the replicated HNC 
approximation \cite{card}, show that $W(q)$
 is minimal at $q=0$,
but has a secondary  minimum at a certain $q=q_{EA}$,
in the temperature range $T_K<T<T_c$. The behaviour around this second,
 metastable, minimum  
corresponds to phenomena that can be observed 
on time scales shorter than the lifetime of the metastable state
(this lifetime is infinite in the mean field models, but becomes finite in
a real system because of nucleation effects).

Let us summarize the present situation in a few words. There 
is a far reaching analogy between structural glasses and the theoretical mean
field models of discontinuous spin glasses. Inspired by this analogy, 
some of the powerful methods and concepts used in spin glass theory have been
applied to standard simple models of glass formers like binary Lennard Jones
systems. These allow to perform some first principle computations of the
equilibrium properties of these systems. I think that some real progress
has been made, but the hardest part is still ahead of us: we now need to 
understand the nucleation properties in order to be able to determine
the lifetime of metastable states and to compute the
fast increase of the relaxation time in the regime between $T_c$ and $T_K$.
The out of equilibrium dynamics also contains many open problems.
The aging behaviour can be found within several different theoretical 
approaches, but these in fact relate to rather different physical mechanisms.
It seems to me rather plausible that a glass, when cooled, will evolve 
through  different aging regimes, starting with a regime (well
described by the mean field dynamics)  where it
seeks unstable directions in the energy landscape, and evolving to a 'trap-like'
regime at longer times. More work, theoretical and
experimental is needed in order to disentangle these
various regimes.
My belief is that a careful study of some recently introduced lattice models
\cite{birmez} could offer the best route to answering all these questions. 

It is a great pleasure to thank Giorgio Parisi for the 
 collaboration which led to the work described in the last part of this paper.



\begin{thebibliography}{99}

\bi{mmaspen}
M. M\'ezard, ``First steps in glass theory'', in {\it More is different},
M.Phuan Ong and Ravin N. Bhatt ed., 
Princeton University Press 2001 (Princeton, New Jersey).

\bi{glass_revue}
Recent reviews can be found in:  C.A. Angell, Science, 
{\bf 267}, 1924 (1995); P.De Benedetti, `Metastable liquids', Princeton 
University
Press (1997); G.Tarjus and D. Kivelson, to appear in "Jamming" volume, A. Lui
and S. Nagel eds.; J.J\"ackle, Rep.Prog.
Phys. {\bf 49} (1986) 171.
\bi{kobrev} W. Kob, cond-mat/9911023.
\bi{youngbook} 
A revue of recent developments, and references to previous work
can be found in  "Spin glasses and random fields",
A.P. Young ed., World Scientific (Singapore) 1998.
\bi{MPV} For a review, see  M. M\'ezard, G. Parisi and M.A. Virasoro,
 {\sl Spin glass theory and beyond}, World Scientific (Singapore 1987)
\bi{kepler} See for instance the recent work on Kepler's conjecture,
www.math.lsa.umich.edu/~hales/countdown/.
\bi{optrev} O.C.Martin, R.Monasson and R.Zecchina, cond-mat/0104428.
\bi{codes} N. Sourlas, cond-mat/0106570.
\bi{mingame} D.Challet, M.Marsili and R.Zecchina, cond-mat/9904392.
\bi{donati}
C.Donati, S. Franz, G.Parisi ans S.C. Glotzer, cond-mat/9905433.

\bi{fn1} Of course there is always an ambiguity due to the existence of the crystal.
In good glass formers, the nucleation time for the formation of the crystal
becomes so huge at the temperatures close to $T_g$ that one can forget about
the existence of the crystal phase: although there is no such phase as an
``equilibrated supercooled liquid phase'' in principle, 
it exists in practice, when the
ratio of crystallisation time to equilibration of the liquid is very large.

\bi{fn2} Sometimes the term 'ideal glass transition' is also
used to characterize the mode coupling temperature. However in many systems 
 we know for sure that there is no transition at the mode coupling temperature,
and I prefer to  use this term as describing the putative transition taking
place at $T_K$.
\bi{RiAn}
  R. Richert and C.A. Angell, J.Chem.Phys. {\bf 108} (1999) 9016.
\bi{kauzmann}
 A.W. Kauzman, Chem.Rev. {\bf 43} (1948) 219.
\bi{AdGibbs}
G. Adams and J.H. Gibbs J.Chem.Phys {\bf 43} (1965) 139; J.H. Gibbs and E.A.
Di 
Marzio, 
J.Chem.Phys. {\bf 28} (1958) 373.
\bi{REM}
B. Derrida, {\em Phys. Rev.}  {\bf B24}, 2613 (1981)
\bi{GrossMez}
D.J. Gross and M. M\'ezard, Nucl. Phys. {\bf B240} (1984) 431.
\bi{KiThWo}
 T.R. Kirkpatrick and P.G. Wolynes,  Phys. Rev. {\bf
A34}, 1045 (1986); T.R. Kirkpatrick and D. Thirumalai, Phys. Rev. Lett. {\bf 
58},
2091 (1987); T.R. Kirkpatrick and D. Thirumalai, Phys. Rev. {\bf B36}, 5388 
(1987); 
T.R. Kirkpatrick, D. Thirumalai and P.G. Wolynes,  Phys. Rev. {\bf
A40}, 1045 (1989).
\bi{crisanti} A. Crisanti, H. Horner and H.J. Sommers,
Z. Physik B {\bf 92}, 257 (1993).
\bi{nodis1}
 J.-P. Bouchaud and M. M\'ezard; J. Physique I (France) {\bf 
4} (1994) 1109.
E.  Marinari, G.  Parisi and F.  Ritort; J.  Phys.  {\bf A27} (1994) 7615; J.  
Phys.  {\bf A27} (1994) 7647.

\bi{nodis2}
P.Chandra, L.B.Ioffe and D.Sherrington, Phys. Rev. lett. {\bf 75} (1995) 713,
and cond-mat/9809417.
P.Chandra, M.V. Feigelman and L.B.Ioffe, Phys. Rev. lett. {\bf 76} (1996) 4805.

\bi{nodis3}
 S. Franz and  J. Hertz, {\it Phys. Rev. Lett.} {\bf 74}, 2114 (1995);
J.P. Bouchaud, L. F. Cugliandolo, J. Kurchan and M. M\'ezard,
{\it Physica} A {\bf 226}, 243 (1996).

\bi{ferropspin}S. Franz, M. Mezard, F. Ricci-Tersenghi, M. Weigt, R. Zecchina,
      Europhys. Lett. {\bf 55} (2001) 465.

\bi{plaq} A. Lipowski and D. Johnston, Phys. Rev. E {\bf
61}, 6375 (2000).  M.R. Swift, H. Bokil, R.D.M. Travasso and
A.J. Bray, Phys. Rev. B {\bf 62}, 11494 (2000).


\bi{gotze}
For a review on mode couping theory,
see W. Gotze, in {\em Liquid, freezing and the Glass transition},
Les
Houches
(1989), J. P. Hansen, D. Levesque, J. Zinn-Justin editors, North Holland.
\bi{MCexp}
Some discussion of the experimental situation can be found in
H. Z. Cummins and  G. Li,
{\it  Phys. Rev.} E {\bf 50}, 1720  (1994);
H. Z. Cummins, W.M. Du, M. Fuchs, W. Gotze, S. Hildebrand, A.
Latz, G. Li and  N.J. Tao,  {\it  Phys. Rev.} {\bf 47}, 4223 (1993);
P. K. Dixon, N. Menon and  S. R. Nagel,
{\it  Phys. Rev.} E {\bf 50}, 1717 (1994).
For an enlightening introduction to the experimental
controversy, see the series of Comments in {\it  Phys. Rev.} E:
X.C. Zeng, D. Kivelson and G. Tarjus,
{\it Phys. Rev.} E {\bf 50}, 1711 (1994).


\bi{cririt}
A. Crisanti and F. Ritort,
cond-mat/9911226 and cond-mat/9911351.

\bi{BCKM}
For a review, see
J.-P.  Bouchaud, L.  Cugliandolo, J.  Kurchan., M M\'e\-zard, in "Spin glasses
and random fields", A.P.Young editor, Worlds Scientific 1998.

\bi{cuku} L.  F.  Cugliandolo and J.Kurchan, Phys.  Rev.  Lett.  {\bf 71}, 
1 (1993).

\bibitem{fdr} S. Franz, M. M\'ezard, G. Parisi and L. Peliti,
Phys. Rev. Lett. {\bf 81} 1758 (1998); {\em The response of glassy systems
 to random perturbations: A bridge between equilibrium and 
off-equilibrium}, cond-mat/9903370, to appear in J.Stat.Phys.

\bi{gpglass}
 G. Parisi Phys.Rev.Lett. {\bf 78}(1997)4581.

\bi{bk1}
 W. Kob and J.-L. Barrat, Phys.Rev.Lett. {\bf 79} (1997) 3660.

\bi{bk2}
 J.-L. Barrat and W. Kob, cond-mat/9806027.
\bi{leo}
R. Di Leonardo, L. Angelani, G. Parisi and G. Ruocco, cond-mat/0001311.

\bibitem{fdr_exp}
T.S. Grigera and N.E. Israeloff, cond-mat/9904351;
L. Bellon, S. Ciliberto, C. Laroche cond-mat/0008160.

\bibitem{jorge}
See J.Kurchan, cond-mat/0110317, and references therein.


\bibitem{crisomtap}
 A.~Crisanti and H.-J. Sommers, J. Phys. I (France) {\bf 5}, 805 (1995);
A.~Crisanti, H.~Horner and H-J~Sommers, {\it Z.Phys.} B  {\bf 92},  257 (1993)

\bibitem{talag} M. Talagrand, Prob. Theor. Rel. Fields {\bf 117} (2000) 303;
A. Bovier  and B. Niederhauser,
cond-mat/0107376.


\bi{IS}
M. Goldstein, J. Chem. Phys. {\bf 51}, 3728 (1969);
F.H. Stillinger and T.A. Weber,Science {\bf 225} (1984) 983;
 F.H. Stillinger, Science {\bf 267} (1995) 1935.
\bi{inherent}
 S. Sastry, P.G. Debenedetti and
F.H. Stillinger, Nature {\bf 393}, 554 (1998);
W. Kob, F. Sciortino and P. Tartaglia, cond-mat/9905090;
 F. Sciortino, W. Kob  and P. Tartaglia, cond-mat/9906278;
S. B\"uchner and A. Heuer, cond-mat/9906280.

\bi{schroder}
T.B. Schroder, S. Sastry, J.C. Dyre and S.C. Glotzer, cond-mat/9901271.

\bi{Still_slope}
F.H. Stillinger, J. Chem. Phys. {\bf 88} (1988) 7818.

\bi{BiMo} G. Biroli and R. Monasson, cond-mat/9912061.

\bi{landscape}T.S. Grigera, A. Cavagna, I. Giardina, G. Parisi,
cond-mat/0107198.

\bi{cav} A. Cavagna, Europhys. Lett. {\bf 53} (2001) 490, cond-mat/9910244.

\bi{pot}  S. Franz ang G. Parisi, J. Physique I {\bf 5} (1995) 1401; 
 Phys.Rev.Lett.  {\bf 79} (1997) 2486.
\bi{card}
 M.Cardenas, S. Franz and G. Parisi, cond-mat/9712099.

\bi{MP_Trieste} This is reviewed in more details in M.M\'ezard and G.Parisi,
proceedings of the Trieste conference on "Unifying Concepts in Glass Physics"
cond-mat/0002128, to appear in J. Phys.

\bi{MePa1} M. M\'ezard and G. Parisi, Phys.  Rev.  Lett.  {\bf 82}, 747 
(1998).
 
\bi{MePa2} M. M\'ezard and G. Parisi J. Chem.  Phys. {\bf 111}, 1076 (1999).
\bi{remi} R. Monasson, {\em Phys. Rev. Lett.} {\bf 75}, 2847 (1995).
\bi{Me} M. M\'ezard, Physica A {\bf 265}, 352 (1999).
\bi{sferesoft} B. Coluzzi, M. M\'ezard, G. Parisi and P. Verrocchio,
J. Chem.  Phys. {\bf 111},9039 (1999).
\bi{latz} A. Latz, cond-mat/0106086.
\bi{LJ} B. Coluzzi, G. Parisi and P. Verrocchio, {\em Lennard-Jones
  binary mixture: a 
thermodynamical approach to glass transition}, cond-mat/9904124.

\bi{LJ2} B. Coluzzi, G. Parisi and P. Verrocchio, {\em The thermodynamical
liquid-glass transition in a Lennard-Jones binary mixture}, 
cond-mat/9906124.

\bi{parisi_c} T.S. Grigera and G. Parisi, Phys. Rev. E 63, 045102(R) (2001).

\bi{krauth} 
L. Santen and W. Krauth, Nature 405, 550-551 (2000).



\bi{birmez} G.Biroli and M. M\'ezard, cond-mat/0106309.
\end{thebibliography}
\end{document}